\documentstyle[tighten,prl,aps]{revtex}
\begin{document}
%
\def\eps{\varepsilon}
\def\hsp{\hspace*{-0.6cm}}
%
%
\draft
\title
{
Remnant Fermi surface in a pseudogap regime of the two-dimensional %
Hubbard model at finite temperature
}
\author{
Taiichiro Saikawa\footnote{
Present address: Dipartimento di Scienze Fisiche ``E.R. Caianiello",
Universit\`{a} degli Studi di Salerno,
84081 Baronissi (Salerno), Italy.}
and Alvaro Ferraz\footnote{
Present address: Institut de physique theorique,Universit\'{e}
de Fribourg, Perolles, CH-1700 Fribourg, Switzerland.}
}
\address{
Laborat\'{o}rio de Supercondutividade,\\
Centro Internacional de F\'{\i}sica da Mat\'{e}ria Condensada,\\
Universidade de Bras\'{\i}lia,\\
CEP 70919-970 Bras\'{\i}lia-DF, Brazil
}

\date{\today}
\maketitle
\begin{abstract}
A precursor effect on the Fermi surface
in the two-dimensional Hubbard model at finite temperatures
near the antiferromagnetic instability
is studied using three different itinerant approaches:
the second order perturbation theory, the paramagnon theory (PT),
and the two-particle self-consistent (TPSC) approach.
In general, at finite temperature, the Fermi surface of the interacting
electron systems is not sharply defined due to the broadening effects
of the self-energy.  In order to take account of those effects
we consider the single-particle spectral function $A({\bf k},0)$
at the Fermi level, to describe the counterpart of the Fermi surface
at $T=0$.
We find that the Fermi surface is destroyed close to
the pseudogap regime due to the spin-fluctuation effects
in both PT and TPSC approaches.
Moreover, the top of the effective valence band is located around
${\bf k}=(\pi/2,\pi/2)$ in agreement with earlier investigations
on the single-hole motion in the antiferromagnetic background.
A crossover behavior from the Fermi-liquid regime to the
pseudogap regime is observed in the electron concentration dependence
of the spectral function and the self-energy.
\end{abstract}

\pacs{
PACS numbers:
}


\section{Introduction}
Recently, investigations of the Fermi surface (FS) have been a crucial
issue in the study on the high-$T_c$ cuprate superconductors.
Angle-resolved photoemission experiments \cite{MARSHALL96,NORMAN98A}
have found evidences of deviations from conventional Fermi liquid
demonstrating the existence of the pseudogap formation above $T_c$
in the single-particle spectral function and the breakdown
of the FS
at the so-called ``hot-spot'' region in the Brillouin zone.

The Fermi surface is an important physical property which characterizes
the low temperature behavior of the normal phase of fermion systems.
If the stability of the Fermi liquid state is verified,
the Luttinger theorem \cite{LUTT} concerning
the conservation of the Fermi surface volume holds.
However, if the interacting fermion system is in a state close to
some instability, the precise definition of both the Fermi liquid and
the FS may become uncertain.

Some indications of the deviations from the conventional Fermi surface
behavior have already been seen in the two-dimensional (2D) Hubbard model.
Quantum Monte Carlo (QMC) simulations \cite{BULUT94c,PREUSS,GROBER99}
indicated the change of the FS topology for small hole dopings
in the strong coupling regime.
This topological change is characterized by the FS closing
around ${\bf k}=(\pi,\pi)$ despite the (small) hole dopings.
In the weak coupling regime of the Hubbard model,
using second order perturbation theory,
Zlati\'c, Schotte, and Schliecker \cite{ZLATIC95} obtained that
the FS changes its topology due to the self-energy effects
for small dopings.
In contradiction with that, within the same framework of
perturbation theory, Halboth and Metzner \cite{HALBOTH97} have shown
that the modification of the FS is very small
for weak coupling at zero temperature.
Thus, even in the weak-coupling regime,
the FS of the Hubbard model leads to controversy.

In this paper, we focus on the modification suffered by the FS
of the 2D Hubbard model near the antiferromagnetic instability
in the weak-coupling regime.
The 2D Hubbard model produces several different electronic phases
according to the choice of the parameters.
For the highly doped case, the system is in the conventional
Fermi liquid regime with a well-defined FS.
On the other hand, in the ground state at half-filling,
the system becomes an antiferromagnetic insulator.  Thus, for
an appropriate set of parameters, the zero temperature phase transition
will take place, and quite probably, the conventional Fermi surface
picture is lost before that.

However the long range antiferromagnetic order at half-filling
is automatically destroyed at sufficiently high temperatures or
finite doping.
Thus, assuming paramagnetic background at finite temperature
we use itinerant approaches such as the second order perturbation
theory, the paramagnon theory\cite{TAI99a} and also the recently developed
two-particle self-consistent approach\cite{VILK94,VILK96,VILK97a}
to calculate the effects produced on the FS.
Varying the electron concentration, the coupling constant and
the temperature, the single-particle spectral function is calculated
in all those approaches.
In general the FS is well defined for a normal metal at $T=0$.
At finite temperature, however, we cannot define the FS
sharply. To describe the counterpart
of the Fermi surface at finite temperatures,
we introduce the spectral function $A({\bf k},0)$
at the Fermi level. The meaning of this quantity is discussed
in detail.

We will see that the FS of the Hubbard model at finite
temperature shows remarkable modifications.
As we approach the antiferromagnetic instability condition,
the FS is destroyed at the $(\pi,0)$ regions in the
Brillouin zone.
The calculated effective band dispersion indicates that
the anisotropic pseudogap and the destruction of the FS
are both manifestations of the precursor of the antiferromagnetic
instability at half-filling.
Besides that, we note that there is a trace of the FS
in the pseudogap regime in which there is no effective band dispersion
crossing the Fermi level.
Also, by varying the electron number per site $n$,
we find that the spectral function and self-energy clearly show
a crossover from the Fermi liquid regime at the smaller $n$
to the pseudogap regime near the half-filling.

\section{Model}
We start with the standard 2D Hubbard Hamiltonian defined by
\begin{equation}
   H = -t \sum_{<i,j>, \sigma} 
          a^{\dagger}_{i \sigma} a_{j \sigma}
      +U \sum_{i} n_{i,\uparrow} n_{i,\downarrow}
      -\mu \sum_{i, \sigma} n_{i, \sigma}
\end{equation}
where $t$ is the nearest-neighbor transfer matrix on the square lattice,
$U$ is the on-site repulsive coupling constant,
$\mu$ the chemical potential
and $n_{i \sigma}= a^{\dagger}_{i \sigma} a_{i \sigma}$.
The Fermi surface can be determined using
the single-particle spectra which is calculated with the full Green's
function.
The spectral function $A({\bf k},\omega)$ is given in terms of
the imaginary part of the retarded Green's function:
$
A({\bf k},\omega) = -(1/\pi){\rm Im}G({\bf k},\omega)
$
where $G({\bf k},\omega)$ is given by
\begin{equation}
G({\bf k},\omega)^{-1} = \omega + i\eta - (\eps_{\bf k}-\mu)
   -\Sigma({\bf k},\omega)
\label{DEQF}
\end{equation}
with $\eps_{\bf k}=-2t (\cos k_x + \cos k_y) $ and
$\Sigma({\bf k},\omega)$ is a proper self-energy which is
given in the following subsection.
The chemical potential $\mu$ is determined to be consistent with
a given electron concentration $n$ which satisfies
\begin{equation}
n = \sum_{{\bf k} \sigma} \int_{-\infty}^{\infty} d\omega
f(\omega) A({\bf k},\omega)
\end{equation}
for fixed values of $U/t$ and $T/t$.
Here $f(\omega)$ is the Fermi distribution function defined
by $f(\omega) = 1/[\exp(\omega/T) + 1]$.

Next, we introduce self-energies for three different approaches;
the paramagnon theory\cite{TAI99a},
the second order perturbation theory, and the two-particle
self-consistent approach\cite{VILK94,VILK96,VILK97a}.

\subsection{Paramagnon theory and second order perturbation theory}
The paramagnon theory (PT) is probably the simplest way to take account
of the magnetic instability in the electronic states qualitatively.
In that approach, the divergence of the spin susceptibility
in the random phase approximation (RPA)
signals the antiferromagnetic instability.
At $T=0$, due to the perfect nesting
condition, the real part of the one-loop polarization bubble
${\rm Re}\chi_0({\bf q},0)$ diverges, and the PT approach
cannot be simply applied.
$\chi_0$ is given by
\begin{equation}
\chi_0({\bf q}, \nu_m) = {\sum_{\bf k}}
   \frac{f(\eps_{{\bf k}+{\bf q}}-\mu_0)-f(\eps_{\bf k}-\mu_0)}
   {i\nu_m -(\eps_{{\bf k}+{\bf q}} - \eps_{\bf k})}
   \label{CHI00}
\end{equation}
with $\nu_m= 2 m \pi T$ and $\mu_0$ is the chemical potential
in the non-interacting case.
At non-zero temperature, however, ${\rm Re}\chi_0({\bf q},0)$ remains
finite even at half-filling. Thus, the RPA spin susceptibility does
not diverge for suitable small values of the coupling constant $U$
and the system remains in its paramagnetic regime.
We perform all calculations in the present paper with suitable choices
of the temperature $T$, the coupling constant $U$,
and the electron concentration $n$.
A zero-temperature version of our approximation has been used 
by Kampf\cite{KAMPF94} to calculate the spectral
function of the doped Hubbard model.
One drawback of the paramagnon theory is that it does predict
the antiferromagnetic instability in two dimensions
violating in this way the Mermin-Wagner theorem.
However, if we avoid this transition regime
by working at finite temperatures,
as we have shown in our previous work\cite{TAI99a},
this approach gives the qualitatively correct
non-Fermi-liquid energy dependence
in the self-energy with the corresponding anisotropic pseudogap
in the single-particle spectral function.
These results are in general agreement with
several other approaches\cite{VILK96,VILK97a,DEISZ96,SCHMA98}.
Moreover, the anisotropy of the pseudogap, which has been
observed also in the QMC simulation by
Creffield et al \cite{CREF95}, was shown in our previous paper.

To derive the self-energy, we use the standard diagram technique
by treating the interaction term perturbatively.
Since the Hartree-Fock term $Un/2$ is nothing but a constant energy
shift, we include it in the chemical potential.
The self-energy in the PT framework can be written as
\begin{eqnarray}
\Sigma^{\rm PT}({\bf k}, \omega_n) &=&
   U^2 T\sum_{\nu_m} \sum_{\bf q}
    \left[
    \frac{1}{2}\chi_c({\bf q},\nu_m)
    +\frac{3}{2}\chi_s({\bf q},\nu_m)
    \right.
\nonumber \\
& & \left.
    -\chi_0({\bf q},\nu_m)
    \right] G_0({\bf k}+{\bf q},\omega_n+\nu_m)
    \label{SIGPARA}
\end{eqnarray}
where $G_0({\bf k},\omega_n$) is the non-interacting Green's function
with $\omega_n=(2n+1)\pi T$.
Here we have introduced the charge susceptibility by
$\chi_c({\bf q},\nu_m) =
    \chi_0({\bf q},\nu_m)/[1+U\chi_0({\bf q},\nu_m)] $
and the spin susceptibility by
$\chi_s({\bf q},\nu_m) =
    \chi_0({\bf q},\nu_m)/[1-U\chi_0({\bf q},\nu_m)] $.
In the angular bracket in Eq. (\ref{SIGPARA}), $\chi_0({\bf q}, \nu_m)$
is subtracted to avoid the double counting of the lowest order diagram.
$\Sigma({\bf k},\omega)$ in Eq. (\ref{DEQF})
is obtained from $\Sigma^{\rm PT}({\bf k},\omega_n)$
by performing an analytic continuation.

In contrast, the self-energy obtained from  conventional second order
perturbation theory is given by
\begin{equation}
\Sigma^{\rm 2nd}({\bf k}, \omega_n) = U^2 T\sum_{\nu_m} \sum_{\bf q}
    \chi_0({\bf q},\nu_m)
    G_0({\bf k}+{\bf q},\omega_n+\nu_m).
    \label{SIG2ND}
\end{equation}
This is equivalent to the lowest order contribution of the self-energy
$\Sigma^{\rm PT}({\bf k}, \omega_n)$ in the paramagnon theory.
Detailed investigation on $\Sigma^{\rm 2nd}({\bf k}, \omega_n)$
of the Hubbard model is seen in references
\onlinecite{ZLATIC95,HALBOTH97,ZLATIC97}.

\subsection{Two-particle self-consistent approach}
The two-particle self-consistent (TPSC) approach has been developed
by Vilk and Tremblay et al\cite{VILK94,VILK96,VILK97a}.
This approach can give both the spin and charge structure factors
in good quantitative agreement with the results
of the quantum Monte Carlo simulations in the Hubbard model \cite{VILK94}.
Detailed argument on this approach
has been given in their review article\cite{VILK97a}.
Following their formulation, the self-energy of
the single electron Green's function in TPSC approach is given by
$$
\Sigma^{\rm TPSC}({\bf k}, \omega_n) =
   \frac{U}{2} T\sum_{\nu_m} \sum_{\bf q}
    \left[
     U_{\rm ch} \chi^{\rm TPSC}_c({\bf q},\nu_m)
    \right.
$$
\begin{equation}
    \left.
    +U_{\rm sp} \chi^{\rm TPSC}_s({\bf q},\nu_m)
    \right] G_0({\bf k}+{\bf q},\omega_n+\nu_m)
    \label{SIGTPSC}
\end{equation}
where
$\chi^{\rm TPSC}_c({\bf q},\nu_m) =
    \chi_0({\bf q},\nu_m)/[1+U_{\rm ch}\chi_0({\bf q},\nu_m)] $
and
$\chi^{\rm TPSC}_s({\bf q},\nu_m) =
    \chi_0({\bf q},\nu_m)/[1-U_{\rm sp}\chi_0({\bf q},\nu_m)] $.
The Hartree-Fock term has been absorbed again in the chemical potential.
Note the difference of the factor $2$ of our definition
of $\chi_0$ here from the one in their original papers.
The renormalized coupling constants $U_{\rm sp}$
and $U_{\rm ch}$ are determined by solving the self-consistent
equations given by
\begin{equation}
   T\sum_{\nu_m} \sum_{\bf q} 2 \chi^{\rm TPSC}_s({\bf q},\nu_m)
 = n -2 \left< n_{\uparrow} n_{\downarrow} \right>,
\label{SC1}
\end{equation}
\begin{equation}
   T\sum_{\nu_m} \sum_{\bf q} 2 \chi^{\rm TPSC}_c({\bf q},\nu_m)
 = n +2 \left< n_{\uparrow} n_{\downarrow} \right> -n^2,
\label{SC2}
\end{equation}
and
\begin{equation}
  U_{\rm sp} = \frac{4U}{n^2} \left< n_{\uparrow} n_{\downarrow} \right>.
\label{SC3}
\end{equation}
Here $\left< n_{\uparrow} n_{\downarrow} \right>$ is the average
double occupancy per site.

Note that contrary to the other two methods the Mermin-Wagner theorem is
satisfied at any finite temperature
in the TPSC approach \cite{VILK94,VILK97a}.
To review it briefly, let's consider the half-filling case.
At half-filling, from Eqs. (\ref{SC3}) and (\ref{SC1}),
we obtain the following self-consistent equation to compute
$U_{\rm sp}$:
\begin{equation}
   T\sum_{\nu_m} \sum_{\bf q} 
   \frac{2 \chi_0({\bf q},\nu_m)}
        {1-U_{\rm sp}\chi_0({\bf q},\nu_m)}
 = 1 - \frac{U_{\rm sp}}{2 U}.
\label{SC}
\end{equation}
If $U_{\rm sp}$ always satisfies the condition
$1>U_{\rm sp} \chi_0({\bf Q},0)$, the magnetic instability is not realized,
namley the Mermin-Wagner theorem holds.
It is convenient to rewrite the above self-consistent equation
in terms of two functions $f_1(u)$ and $f_2(u)$
defined by
\begin{equation}
   f_1(u) = T\sum_{\nu_m} \sum_{\bf q} 
   \frac{2 \chi_0({\bf q},\nu_m)}
        {1-u \chi_0({\bf q},\nu_m)}
\label{f1}
\end{equation}
and
\begin{equation}
  f_2(u) = 1 - \frac{u}{2 U}.
\label{f2}
\end{equation}
The self-consistent equation (\ref{SC}) is equivalent to
the relation $f_1(U_{\rm sp})=f_2(U_{\rm sp})$.
Fig.\ \ref{F1F2} shows a schematic plot of $f_1(u)$ and $f_2(u)$
as the functions of $u$ for finite $T$.
The behavior of $f_2(u)$ is trivial. Note that the slope of $f_2(u)$
is in proportion with $1/U$.
Using the sum rule of $\chi_0$ given in Appendix, we see $f_1(u=0) = 0.5$.
$f_1(u)$ diverges as $u$ approaches $1/\chi_0({\bf Q},0)$.
This divergence was proved by Vilk and Tremblay \cite{VILK97a}.
The intersection of $f_1(u)$ and $f_2(u)$ gives the solution of
the self-consistent equation.  From the plot, we see that
the intersection point always stays within the range
$0<u<1/\chi_0({\bf Q},0)$ for arbitrary $U$.
Namely, the solution of the TPSC self-consistent equation never violates
the Mermin-Wagner theorem.

\section{Fermi surface at finite temperature}
Let's consider how to obtain the information on the Fermi surface
from the full Green's function given by Eq.\ (\ref{DEQF})
at finite temperature.
So far, several different methods to investigate
the FS have been used in the literature.
One straightforward way to define the FS
is given by the quasiparticle poles of the Green's
function through the solutions of
$\eps_{{\bf k}_F} - \mu + {\rm Re}\Sigma({\bf k}_F,0)=0$.
Another way is to use the electron momentum distribution
function $n({\bf k})$ defined by
$n({\bf k})=\int_{-\infty}^{\infty} d\omega f(\omega) A({\bf k},\omega)$.
In this case, the FS is characterized by
each ${\bf k}$ point
associated with a finite discontinuity of $n({\bf k})$.
Those definitions give a proper FS at zero temperature.
However, at finite temperature, since the imaginary part of
the self-energy has a finite value with a non-trivial
${\bf k}$ dependence in general, these definitions are not applicable.
Another definition is to use $n({\bf k}_F)=1/2$.
This is not useful to apply for the pseudogap regime
since this condition is always satisfied at half-filling
and at finite temperature.
Maximum slope of $\nabla_{\bf k} n({\bf k})$ is also used in the
determination of the FS obtained with the angle-resolved photoemission
\cite{ARPES1,ARPES2}.
However, because $n({\bf k})$ is obtained from a quantity
which is integrated
over the energy, it has a maximum slope even if a gap or a pseudogap
exists.
At finite $T$ the FS is often
defined approximately by the ${\bf k}$ points at which the effective
band dispersion crosses on the Fermi level.  The effective
band dispersion is obtained from the maximum intensity points
of the spectral function.

Strictly speaking, at finite temperature we cannot define a sharp
FS because of the broadening effects in the spectral
function.
However, in principle, the Fermi level can be calculated for any sets
of parameters of the system if the number of electrons is known.
We therefore use this route to establish the Fermi surface.
Taking into account the above consideration on the FS,
instead of employing conventional definitions,
we consider the electron distribution at the Fermi level.
Such distribution can be represented by a single-particle spectral
function written as
\begin{equation}
A({\bf k},0) =
 \frac{1}{\pi}
 \frac{ \mid {\rm Im}\Sigma({\bf k},0) \mid }
 {[ \eps_{\bf k}-\mu-{\rm Re}\Sigma({\bf k},0)]^2
 +[ {\rm Im}\Sigma({\bf k},0)]^2}.
 \label{AK0}
\end{equation}
It is evident that it satisfies a sum-rule
$N(0) = \sum_{\bf k} A({\bf k},0)$ where $N(0)$ is
the density of states (DOS) at the Fermi level.
$A({\bf k},0)$ contains all the information of the single-particle
states at the Fermi level and possesses enough information to study
the properties of the Fermi surface.

We easily see that $A({\bf k},0)$
has the following features, which covers the basic properties of
the conventional FS.
When the system is described by the conventional Fermi-liquid
at zero temperature, we have that
\begin{equation}
A({\bf k},0) = \delta( \eps_{\bf k}-\mu-{\rm Re}\Sigma({\bf k},0) )
\end{equation}
and the peak points of the delta-function trace
a sharply defined FS in the ${\bf k}$ space.
Thus, $A({\bf k},0)$ recovers the conventional FS at the ground states.
In a Fermi liquid at finite temperature,
the spectral function becomes broad around the Fermi level.
Quite naturally, $A({\bf k},0)$ covers this feature because
it has a finite width and it draws a broad trace in the ${\bf k}$ space.
On the other hand, when the system does not have a spectral weight
at all at the Fermi level ($N(0)=0$) -- for instance,
an insulating state with a real gap -- $A({\bf k},0)$ is identically
zero for all wave vectors.
Thus, $A({\bf k},0)$ represents all cases ranging from the well-defined
FS to the vanishing FS limit.
As a result, $A({\bf k},0)$ can be used to 
trace the evolution of the FS in all cases.
Observed Fermi surface in angle-resolved photoemission experiments
is extracted from the finite weight of the single particle spectra
in the vicinity of the Fermi level.  Here too, in this respect,
it is meaningful to introduce $A({\bf k},0)$
in order to study the FS structure.

As we see from the argument above, the generalized FS
at finite temperatures is no longer strictly speaking a {\em surface}
in the momentum space. It is not clear if it could be used as
the counterpart of the $T=0$ FS.  This issue
could be solved by
analogy with what happens to the band dispersion in both
non-interacting and interacting systems.
The electronic states in the free-fermion system in a lattice
is completely described by a non-interacting band dispersion
$\omega=E_{\bf k}-\mu$.  In the interacting system, within certain
theoretical treatment of the interaction term, the spectral
function $A({\bf k},\omega)$ contains all the information of
the single-particle states, but the (effective) band dispersion
cannot fully describe the electronic states.
The FS of the non-interacting system is given
by $E_{\bf k}-\mu = 0$.  On the other hand, $A({\bf k},0)$
at the Fermi level continues to carry all necessary information
concerning the electronic states.
As a result it replaces the band dispersion in the description of
the electronic states in the interacting system.
In the following section of this paper, we use the term, Fermi surface,
in this extended sense.

\section{Results}
Using the self-energies and the Green's functions presented
in the previous
section, numerical calculations were performed in the same manner
as in our previous work\cite{TAI99a}.
A mesh of about 4000 grids in the first Brillouin zone has been used
to compute the momentum integral to obtain the imaginary parts
of both charge and spin susceptibilities.
A $120 \times 120$ mesh on the Brillouin zone has been used
to perform the momentum integrations for the self-energies
to obtain the Green's functions and the spectral functions.

\subsection{Second order perturbation theory}
In Fig.\ \ref{DOS2ND} we show the temperature dependence of DOS
calculated using the second order perturbation theory for $U=4.0t$
and $n=1.0$.
Detailed analyses of the self-energy and the single-particle spectra
at finite temperatures and finite dopings
have already been done
by Zlati\'c, Grabowski, and Entel \cite{ZLATIC97}.
A main temperature dependence in DOS appears only around
the Fermi level ($\omega=0$). We do not see any pseudogap-like structure
in DOS in this approximation.
At $T=0.5t$, the maximum of DOS at the Fermi level is broad.
For a lower temperature $T=0.1t$, this maximum becomes a sharp peak.

Figure \ref{FS2ND} shows the gray-scale density plot of $A({\bf k},0)$
for the corresponding parameters used in Fig.\ \ref{DOS2ND}.
As emphasized earlier on, the density plot of
$A({\bf k},0)$ in Fig.\ \ref{FS2ND} gives a counterpart of the
Fermi surface. In this plot, the larger (smaller or zero) intensity
of $A({\bf k},0)$ is represented by the bright (dark) region and
the axes for both $k_x$ and $k_y$ directions are normalized by $\pi$.
For $T=0.1t$ (lower panel), a well-defined FS has been
obtained.
The broadening in the plot for $T=0.5t$ (upper panel) can be
understood as the temperature effect in the self-energy.
Thus, within our results, the second order perturbation theory
gives the conventional Fermi-liquid-like behavior both in the
single-particle spectra and in the FS.
Namely, this approach does not explain the breakdown
of the Fermi surface.
Our calculation has been done at half-filling only
whereas the calculation of Zlatic et al\cite{ZLATIC97} is
for large doping ($n=0.8$). They found a topology change
of FS, but their choice of parameters may not be entirely
suitable for that regime.

\subsection{Paramagnon theory}
Figure \ref{FSN} shows the $n$-dependence of $A({\bf k},0)$ for $U=2.0t$
at $T=0.22t$ calculated in terms of $\Sigma^{\rm PT}({\bf k}, \omega_n)$.
For $n=0.8$ (upper panel), a clear and large FS
is obtained indicating that the system is in the conventional
Fermi liquid phase.
For $n=0.9$ (middle panel), the FS volume is enlarged by
the increment of the electron concentration. In addition, the distribution
around $(\pi,0)$ becomes broader.
For $n=1.0$ (lower panel), the FS around $(\pi,0)$
regions is completely destroyed and there remains well-defined Fermi
surface segments only around $(\pi/2,\pi/2)$ regions.
This anisotropic destruction of the FS is caused by
the anisotropy of the pseudogap formation.
The origin of this kind of anisotropic pseudogap has been
explained in our previous work\cite{TAI99a}.
The strong low-energy antiferromagnetic Stoner enhancement in the spin
susceptibility produces the non-Fermi-liquid energy dependence in
the fermion self-energy.  The momentum dependence of the self-energy
is determined by the anisotropy of the band flatness of the non-interacting
band dispersion at the Fermi level.
As a result of those effects, the pseudogap opens first at the $(\pi,0)$
regions, and the FS is partially destroyed.

In Fig.\ \ref{FSU} we plot the $U$-dependence of the spectral function
$A({\bf k},0)$ within the paramagnon theory
for $T=0.5t$ at half-filling.
For $U=2.0t$ (upper panel), the FS is
rather sharply defined as for the non-interacting case.
As $U$ becomes large, the distribution becomes broad and
the FS cannot be sharply defined.
The tendency for broadening is remarkable in the $(\pi,0)$ regions.
The original structure of the spectral function around $(\pi,0)$
is almost washed out for $U=3.0t$.
Instead, there is a hollow structure centered on $(\pi,0)$.
The $(\pi/2,\pi/2)$ regions have relatively large intensities.

In Fig.\ \ref{DISPPT} we show the effective band dispersion
obtained from the main peaks of the spectral functions
$A({\bf k}, \omega)$ for the same parameter as the lower panel
in Fig.\ \ref{FSU}.
The pseudogap around $(\pi,0)$ is larger than the one
at $(\pi/2,\pi/2)$,
and segments of the shadow bands\cite{KAMPF90} can be
seen around this point.
This result indicates that the top of the effective lower (valence)
band is located at ${\bf k}=(\pi/2,\pi/2)$.
This result is consistent with the earlier investigations
of the single-hole motion in the 2D antiferromagnetic
background\cite{CHUB98}.
In Fig.\ \ref{DISPPT} it is important to realize that the effective band
does not cross the Fermi level. In conventional meaning we cannot define
the FS in this regime, but the spectral weight remains
at the Fermi level and it forms a very broad remnant Fermi surface
structure in the momentum space as we see in the lower-most
panel of Fig.\ \ref{FSU}.

\subsection{Two-particle self-consistent approach}
Figure \ref{TPSCFS} shows the plot of $A({\bf k},0)$ in the TPSC
for $U=4.0t$ at half-filling for $T=0.5t$ (upper panel)
and $T=0.3t$ (lower panel).
Despite the above mentioned difference between the TPSC and PT,
both approaches show qualitetive agreement with respect to the
destruction of FS.
The quantitative diffference between the two methods is due
to the fact that in the PT case the interaction effects are
overestimated by the lack of vertex corrections. 

In Fig.\ \ref{TPSCDOS} we show the temperature dependence of
DOS $N(\omega)$ in the TPSC approach for $U=4t$ and $n=1$.
As we lower the temperature, the gap-like structure grows
at the Fermi energy.  That behavior is well correlated
with the destruction of the FS observed
in Fig.\ \ref{TPSCFS}.

The electron concentration dependence of the spectral intensity
$A({\bf k},0)$
is shown in Fig.\ \ref{TPSCFS2} for $U=4t$ and $T=0.25t$.
For $n=0.85$ (upper panel), the FS is still defined
well. At $n=0.9$, the FS enlarged but in the regions
near the Brillouin zone it gets broader.
At $n=0.95$, the FS region near $(\pi,0)$ becomes
very unclear, while the other FS sectors remain
relatively well-defined. Again, this indicates the anisotropic breakdown
of the Fermi liquid picture on the Fermi level.

In Fig.\ \ref{TPSCSPC} we plot the spectral functions and
the self-energies for various electron concentrations
for $T=0.25t$ and $U=4.0t$.
For convenience, the momentum ${\bf k}_F$ has been assumed
to be the same one in the $(1,0)$ direction
on the non-interacting Fermi surfcae for each $n$.
At $n=0.8$, the spectral function (a) has a single peak
at the Fermi level. This peak corresponds to the conventional
Fermi liquid quasiparticle.
As increasing $n$ towards half-filling, the peak becomes broader
and shifts into the $\omega<0$ side. On the other hand, a shoulder
appears in the $\omega>0$ side in the spectral function ($n=0.9$)
and in the end this turns into another peak.  Thus the quasiparticle
is destroyed, and two-peak structure emerges at $n=1$.
The self-energies also show the breakdown of the Fermi liquid picture.
In our standard knowledge on the Fermi liquid theory,
the real part of the self-energy ${\rm Re}\Sigma$ is linear in $\omega$
and the slope is negative, and the imaginary part of the self-energy
${\rm Im}\Sigma \propto \omega^2$.
At $n=0.8$, the real (b) and imaginary (c) parts of the self-energy
show those Fermi liquid like behaviors.
As $n$ becomes larger, the self-energy dramatically changes
its $\omega$ dependence and indicates deviations
from the Fermi liquid nature.
There are two remarkable features at $n=1$:
(i) the $\omega$ dependence of the real part of the self-energy
resembles to the spin-density-wave self-energy ($\sim 1/\omega$),
(ii) the large negative peak appears
in the imaginary part of the self-energy.

\section{Discussion and Conclusion}
We find that the second order perturbation theory
fails to show the breakdown of the Fermi surface.
The decisive difference between the second order perturbation
and the other two approaches is the fact that 
in the former case the Stoner factor
is not taken into account in the self-energy. We can immediately see this
by comparing the definitions of these self-energies.
As we have already mentioned, the Stoner enhancement is important
to produce the anomalous energy dependence in the self-energy.
If it is combined with the flat-band dispersion
the anisotropic pseudogap is formed in the spectral function.
Because of this difference in self-energy,
the second order perturbation theory FS
is well defined and does not show any anomaly even for dopings
close to half-filling.

As mentioned in Section 1 there has been some controversy
concerning the Fermi surface topology in the 2nd order perturbation
theory framework\cite{ZLATIC95,HALBOTH97,ZLATIC97}.
This issue was recently discussed by Halboth and Metzner (HM) in
\cite{HALBOTH97}. We just want to add that in our calculation
we pay attention to the existence of different values of chemical
potentials for the non-interacting Green's function and
the interacting Green's functions which is renormalized
by the self-energy effects. If one uses a fixed value of chemical
potential in both cases the conservation of electron
number is automatically destroyed. This will in turn produce
discrepancies for any finite doping. This does not seem to
be taken into account in Refs. \cite{ZLATIC95} and \cite{ZLATIC97}.

We discuss next the possibility of the small hole pocket formation
in the 2D Hubbard model.
Langer {\it et al} \cite{LANGER95} and Schmalian{\it et al} \cite{SCHMA96},
using the fluctuation exchange (FLEX) approximation for
the Hubbard model,
have investigated how the low-energy electronic states and
the FS evolve at finite dopings.
They explained\cite{SCHMA96} one scenario of the formation
of the small hole pockets produced by the transfer of
the spectral weight from the main band to the shadow band.
In the range of the parameters used in our calculation,
we have not obtained any sign of the hole pockets
or even the ``shadow'' of the Fermi surface \cite{LANGER95},
while we observed the segments of the shadow bands in
the effective band dispersion.
According to the argument in Ref.\ \onlinecite{LANGER95},
the origin of the shadow band is a large imaginary part
of the self-energy at the ${\bf k}+{\bf Q}$ regions rather than
another pole in the single-particle Green's function.
In their result the imaginary part of the self-energy has
a strong momentum dependence.  That feature has been seen also
in our previous calculation\cite{TAI99a}.
However, the momentum dependence of ${\rm Im}\Sigma({\bf k},\omega)$
produces the anisotropy of the pseudogap in $A({\bf k},\omega)$.
In both PT and TPSC approach, the origin of the double-peak
is clearly associated with the real poles of the full Green's function.
Because of this difference, it is natural that our data
showed neither the shadow of the FS nor the hole-pockets.

Chubukov and Morr\cite{CHUB97} have shown the small hole pocket
formation in an analytical investigation based on
the nearly antiferromagnetic Fermi liquid approach for
the spin-fermion model.
Schmalian {\it et al} \cite{SCHMA98} have also obtained the small
half-pocket-like segments in the FS
for the spin-fermion model.
Although both the PT and TPSC approaches are in a different approximation
level from their semi-phenomenological model, the importance of the
spin-fluctuation is common in both approaches.
To compare with their results we need to study the $t$-$t'$-$U$
Hubbard model\cite{DUFFY95}
which has the same non-interacting band dispersion
including the next-nearest-neighbor hopping $t'$
and to use the Fermi surface topology as they used.
If the same mechanism discussed in the present paper
works in the $t$-$t'$-$U$ model,
we can expect to obtain a half closed small
FS around the $(\pi/2,\pi/2)$ point.

The single-particle spectrum of the $t$-$J$ model is equivalent to 
that of the Hubbard model in the strong coupling limit
if we take no consideration of the upper Hubbard band effect.
Recently, Dai and Su \cite{DAI98} obtained the destruction
of the FS of the $t$-$J$ model, and the FS
for a hole doping clearly shows a topological change of the Fermi
surface which is consistent with the results of the QMC simulations
in the large-coupling Hubbard model\cite{BULUT94c,PREUSS,GROBER99}.
Putikka, Luchini, and Singh\cite{PUTIKKA98} also showed
the violation of the Luttinger theorem in the $t$-$J$ model.

In the present work which deals with the weak-coupling Hubbard model,
we cannot make any decisive statement concerning the validity of
the Luttinger theorem although the complete FS structure
no longer exists.
At half-filling in the arbitrary coupling, the Hubbard model acquires
the particle-hole symmetry.
Hence, presumably the half-filling Fermi surface (if it exists) cannot
show any anomalous topology. As it is well known, in the large
coupling regime, the doped Hubbard model have a preformed gap associated
with the Mott-Hubbard transition.  The energy scale of the preformed gap
is of the order of $U$.
The QMC simulations have shown that the doped hole forms a narrow
quasiparticle band dispersion
on the top of the lower Hubbard band\cite{BULUT94c,PREUSS,GROBER99}.
Namely, there is a jump of the Fermi level
between the infinitesimal hole and electron dopings.
This discontinuity of the Fermi level produces the deviation
of the electron-hole symmetry in the electronic states.
In the weak-coupling regime, since the effect of the preformed gap
at finite temperatures is considerably weak,
the electronic states must preserve the electron-hole symmetry
for an infinitesimal doping, and the topological change cannot
be seen.
Thus, the topology change of the FS
in the large coupling regime may be a manifestation of the effective
violation of the electron-hole symmetry due to the large preformed gap.

In conclusion, we studied the electron distribution $A({\bf k},0)$
at the Fermi level for the 2D Hubbard model in the weak-coupling regime.
This quantity is useful for the generalization of the conventional
Fermi surface concept and can be applied
in all circumstances for all ranges of parameters of the Hubbard model
at finite temperature.
We examined three different itinerant approaches, and
found that the second order perturbation theory fails to
describe the breakdown of the Fermi surface.
In the pseudogap regime, we found that the FS is partially
destroyed in PT and TPSC.
From the anisotropic pseudogap and also from the destruction of the
FS, we observe that the top of the effective valence band
is located at ${\bf k}=(\pi/2,\pi/2)$.
This is in general agreement with prevoius theoretical studies
for a strongly coupled regime\cite{CHUB98}.
We also investigated the $\omega$ dependence of the spectral function
and the self-energy on the FS.
We showed that there exists only quasiparticle like state for small $n$.
In contrast, as we increase $n$, we observe a crossover between
a Fermi-liquid phase and a spin-density-wave-like regime
with the evolution of the precursor pseudogap.

\acknowledgements
We would like to thank Dr. Z. Y. Weng for stimulating discussions.
We also thank Professor H. Kaga for bringing our attention
to the references \onlinecite{VILK96,VILK97a} and for 
very useful conversations.
This work was supported by the Conselho Nacional de
Desenvolvimento Cient\'{\i}fico e Tecnol\'ogico - CNPq and
by the Financiadora de Estudos e Projetos - FINEP.
Most of the numerical calculations were performed with
the supercomputing system at the Institute for Materials Research,
Tohoku University, Japan.

\appendix

\section*{A sum rule for $\chi_0({\bf q},\nu_m)$}

In this appendix, we introduce a sum rule for $\chi_0({\bf q},\nu_m)$
given by
\begin{equation}
   T\sum_{\nu_m} \sum_{\bf q} \chi_0({\bf q},\nu_m)
 = \frac{n}{2} \left( 1-\frac{n}{2} \right).
\end{equation}
This holds for arbitrary electron concentrations and for any temperature.
It can be proved by calling the definition of $\chi_0({\bf q},\nu_m)$
given by Eq.\ (\ref{CHI00}).
Physically speaking, this sum rule states that the total summation
of the particle-hole bubble $\chi_0({\bf q},\nu_m)$ over ${\bf q}$
(in the Brillouin zone) and $\nu_m$ is equal to the products of
the electron concentration $n/2$ and the hole concentration $1-(n/2)$
per spin. It is possible to generalize this sum rule for
the one-loop diagram given in terms of the full Green's functions
$G({\bf k}, \omega_n)$.
In this case, the sum rule becomes
\begin{equation}
   T\sum_{\nu_m} \sum_{\bf q} \chi({\bf q},\nu_m)
 = \frac{n}{2} \left( 1-\frac{n}{2} \right)
\end{equation}
where
\begin{equation}
   \chi({\bf q},\nu_m) =  -T\sum_{\omega_n} \sum_{\bf k}
  G({\bf k}+{\bf q}, \omega_n + \nu_m) G({\bf k}, \omega_n).
\end{equation}
It can be proved in terms of
\begin{equation}
  G({\bf k}, \omega_n) = \int^{\infty}_{-\infty} \frac{d \omega'}{\pi}
  \frac{{\rm Im}G({\bf k},\omega')} {\omega'- i\omega_n}.
\end{equation}


\begin{figure}[h]
\caption{
A schematic plot of the functions $f_1(u)$ and $f_2(u)$.
The solid circles A and B indicate the solutions of the equation
$f_1(u)=f_2(u)$ for larger $U$ and smaller $U$, respectively.
} 
\label{F1F2}
\end{figure}

\begin{figure}[h]
\caption{
Temperature dependence of the density of states $N(\omega)$
within the second order perturbation theory
for $U=4.0t$ at half-filling ($n=1$).
} 
\label{DOS2ND}
\end{figure}

\begin{figure}[h]
\caption{
Spectral intensity $A({\bf k},0)$ in the Brillouin zone
calculated within the second order perturbation
for $U=4.0t$ at half-filling ($n=1$).  Upper panel for $T=0.5t$ and
lower panel for $T=0.1t$.
} 
\label{FS2ND}
\end{figure}

\begin{figure}[h]
\caption{
Spectral intensity $A({\bf k},0)$ in the Brillouin zone
within the paramagnon theory calculated for various electron
concentration $n$ for $U=2.0t$ at $T=0.22t$.
The upper panel is for $n=0.8$, the middle panel for $n=0.9$,
and the bottom panel for $n=1.0$.
} 
\label{FSN}
\end{figure}

\begin{figure}[h]
\caption{
Spectral intensity $A({\bf k},0)$ in the Brillouin zone
within the paramagnon theory calculated for various values of $U$
for $T=0.5$ at half-filling.
From top to bottom, $U=2.0t$, $2.5t$ and $3.0t$, respectively.
} 
\label{FSU}
\end{figure}

\begin{figure}[h]
\caption{
Effective band dispersion obtained from the spectral function
$A({\bf k},\omega)$ within the paramagnon theory calculated
for $U=3.0t$, $T=0.5t$ at half-filling.
} 
\label{DISPPT}
\end{figure}

\begin{figure}[h]
\caption{
Temperature dependence of the spectral intensity $A({\bf k},0)$
in the Brillouin zone
calculated within the two-particle self-consistent approach
for $U=4.0t$ and $n=1$.  Upper panel for $T=0.5t$ and
lower panel for $T=0.3t$.
} 
\label{TPSCFS}
\end{figure}

\begin{figure}[h]
\caption{
Temperature dependence of the density of states
calculated within the two-particle self-consistent approach
for $T=0.3t$ (solid curve) and $0.5t$ (broken curve)
at $U=4t$ and $n=1$.
} 
\label{TPSCDOS}
\end{figure}

\begin{figure}[h]
\caption{
Electron concentration $n$ dependence
of the spectral intensity $A({\bf k},0)$ in the Brillouin zone
calculated within the two-particle self-consistent approach
for $U=4.0t$ and $T=0.25$.
The upper panel is for $n=0.85$, the middle panel for $n=0.9$,
and the lower panel for $n=0.95$.
} 
\label{TPSCFS2}
\end{figure}

\begin{figure}[h]
\caption{
The spectral function (a), the real part (b) and imaginary part (c)
of the self-energy within the two-particle self-consistent approach
at $U=4t$ and $T=0.25t$ for various electron concentration $n$.
The Fermi momentum in the $(1,0)$-direction
on the non-interacting Fermi surface has been used as
${\bf k}_F$ for each $n$.
} 
\label{TPSCSPC}
\end{figure}

\end{document}